\documentclass[%
 twocolumn,
superscriptaddress,
 amsmath,amssymb,
 aps,
]{revtex4-2}

\usepackage{graphicx}
\usepackage{dcolumn}
\usepackage{bm}
\usepackage{xcolor}
\newcolumntype{.}{D{.}{.}{-1}}
\newcolumntype{d}[1]{D{.}{.}{#1}}
\newcommand*{\wn}{cm$^{-1}$}
\newcommand*{\hsm}{H$_{2}$S}
\newcommand*{\Hm}{H$_{2}$}
\newcommand*{\X}{X$^1\Sigma_g^+$}
\newcommand*{\EF}{EF$^1\Sigma_g^+$}
\newcommand*{\F}{F$^1\Sigma_g^+$}
\newcommand*{\GK}{GK$^1\Sigma_g^+$}
\newcommand*{\HH}{H$\bar{\rm H}^1\Sigma_g^+$}

\begin{document}

\title{Photolysis production and spectroscopic investigation of the highest vibrational states in H$_2$ (X$^1\Sigma_g^+$ $v=13,14$)}

\thanks{Published as part of The Journal of Physical Chemistry virtual special issue “Cheuk-Yiu Ng Festschrift”.}%

\author{K.-F. Lai}
 \affiliation{Department of Physics and Astronomy, LaserLaB, Vrije Universiteit \\
 De Boelelaan 1081, 1081 HV Amsterdam, The Netherlands}
 \author{M. Beyer}
 \affiliation{Department of Physics and Astronomy, LaserLaB, Vrije Universiteit \\
 De Boelelaan 1081, 1081 HV Amsterdam, The Netherlands}
 \author{E. J. Salumbides}%
  \affiliation{Department of Physics and Astronomy, LaserLaB, Vrije Universiteit \\
 De Boelelaan 1081, 1081 HV Amsterdam, The Netherlands}
\author{W. Ubachs}%
 \affiliation{Department of Physics and Astronomy, LaserLaB, Vrije Universiteit \\
 De Boelelaan 1081, 1081 HV Amsterdam, The Netherlands}

\date{\today}

\begin{abstract}
Rovibrational quantum states in the \X\ electronic ground state of H$_2$ are prepared in the $v=13$ vibrational level up to its highest bound rotational level $J=7$, and in the highest bound vibrational level $v=14$ (for $J=1$) by two-photon photolysis of H$_2$S.
These states are laser-excited in a subsequent two-photon scheme into \F\ outer well states, where the assignment of the highest ($v,J$) states is derived from a comparison of experimentally known levels in \F, combined with \emph{ab initio} calculations of \X\ levels.
The assignments are further verified by excitation of \F\ population into autoionizing continuum resonances which are compared with multi-channel quantum defect calculations.
Precision spectroscopic measurements of the F-X intervals form a test for the \emph{ab initio} calculations of ground state levels at high vibrational quantum numbers and large internuclear separations, for which agreement is found.
\end{abstract}

\maketitle

\section{Introduction}

The hydrogen molecule has been the benchmark species of molecular spectroscopy since the first analysis of its dipole-allowed absorption spectrum, now over a century ago~\cite{Lyman1906}. Over decades further detailed experiments on the electronic spectrum were performed~\cite{Herzberg1959,Herzberg1972,Chupka1987}, while also the measurements of forbidden vibrational transitions were explored~\cite{Herzberg1949}.
Alongside, and stimulated by experimental observations, the quantum theory of the ground state of smallest neutral molecule was developed with major contributions from James and Coolidge~\cite{James1933}, Ko\l{}os and Wolniewicz~\cite{Kolos1968}, and Wolniewicz~\cite{Wolniewicz1995}. The excited states and the strong effects of non-adiabatic interactions were investigated by Dressler and coworkers~\cite{Dressler1986}.
The theoretical program of refined calculations of the ground state structure was further extended by Pachucki and coworkers including effects of non-adiabatic, relativistic and quantum-electrodynamical (QED) effects~\cite{Komasa2011,Komasa2019}, which has now produced an on-line program (H2SPECTRE~\cite{SPECTRE2019}) to compute level energies for all rovibrational states of the hydrogen isotopomers~\cite{Czachorowski2018}.

Spectroscopic studies of molecular hydrogen have included the behavior at long internuclear separation. Characteristic for the level structure of \Hm\ is the occurrence of double-well potential energy curves for excited states, induced by strong non-adiabatic interactions in this light molecule. In the manifold of $g$-symmetry the lowest of these is the \EF\ state, for which the outer well was investigated~\cite{Marinero1983,Dickenson2012a}, followed by the \GK\ and \HH\ states~\cite{Wolniewicz1985,Reinhold1997}. Similarly double-well states of $^1\Pi_u$ symmetry~\cite{Yu1994,Reinhold1998} and of $^1\Sigma_u^+$ symmetry were investigated~\cite{Kolos1976a,Lange2001}. These studies on long-range effects in the hydrogen molecule were extended to higher energies, leading to observation of exotic phenomena as ion-pair or heavy Rydberg states~\cite{Reinhold2005}, quasi-bound states~\cite{Beyer2016} and shape resonances~\cite{Beyer2018} in the molecular ion.

Various approaches have been followed to investigate \Hm\ in vibrationally excited states of the \X\ electronic ground state, also exhibiting wave function density at large internuclear separation.
Moderately excited $v$-levels were probed in chemical reaction dynamical studies~\cite{Aker1989,Kliner1991}, with hot filaments~\cite{Robie1990,Pomerantz2004}. and in a high voltage discharge~\cite{Dickenson2012a}. Instead of producing the vibrationally excited states over a wide population distribution, Zare and coworkers proposed Stark-induced Raman passage to prepare a single desired state of \Hm~\cite{Mukherjee2017} and recently showed controlled transfer of large population to $v = 7, J = 0$~\cite{Perreault2020}.

Steadman and Baer investigated the production of vibrationally excited states via the two-photon ultraviolet photolysis of \hsm~\cite{Steadman1989}. The results of this one-laser experiment was further investigated in two-laser~\cite{Niu2015b} and three-laser experiments~\cite{Trivikram2016,Trivikram2019} leading to accurate level energies and test of quantum electrodynamics in \X, $v=11-12$.
Alternatively, the UV-photolysis of formaldehyde (H$_2$CO) was used for the production and investigation of \Hm\ in \X, $v=3-9$~\cite{Mitchell2020}.
In these studies the wave function density at large internuclear separation, as occurring in high-$v$ states, was probed via two-photon excitation to the \F\ outer well state.

In the present study, the two-photon UV photolysis production of vibrationally excited \Hm\ from \hsm\ is extended by increasing the photon energy of the dissociation laser. By this means the dissociation channel for producing \X, $v=13-14$ becomes energetically possible. These highly excited vibrations are interrogated with Doppler-free 2+1' resonance multiphoton ionization (REMPI) spectroscopy in a three-laser scheme. Precision measurements probing \F, $v = 0, 1$ outer well levels allow for testing high-accuracy quantum chemical calculations of \Hm\ in the regime of large internuclear separation.

\section{Experiment}

\begin{figure}[b]
\begin{center}
\includegraphics[width=\linewidth]{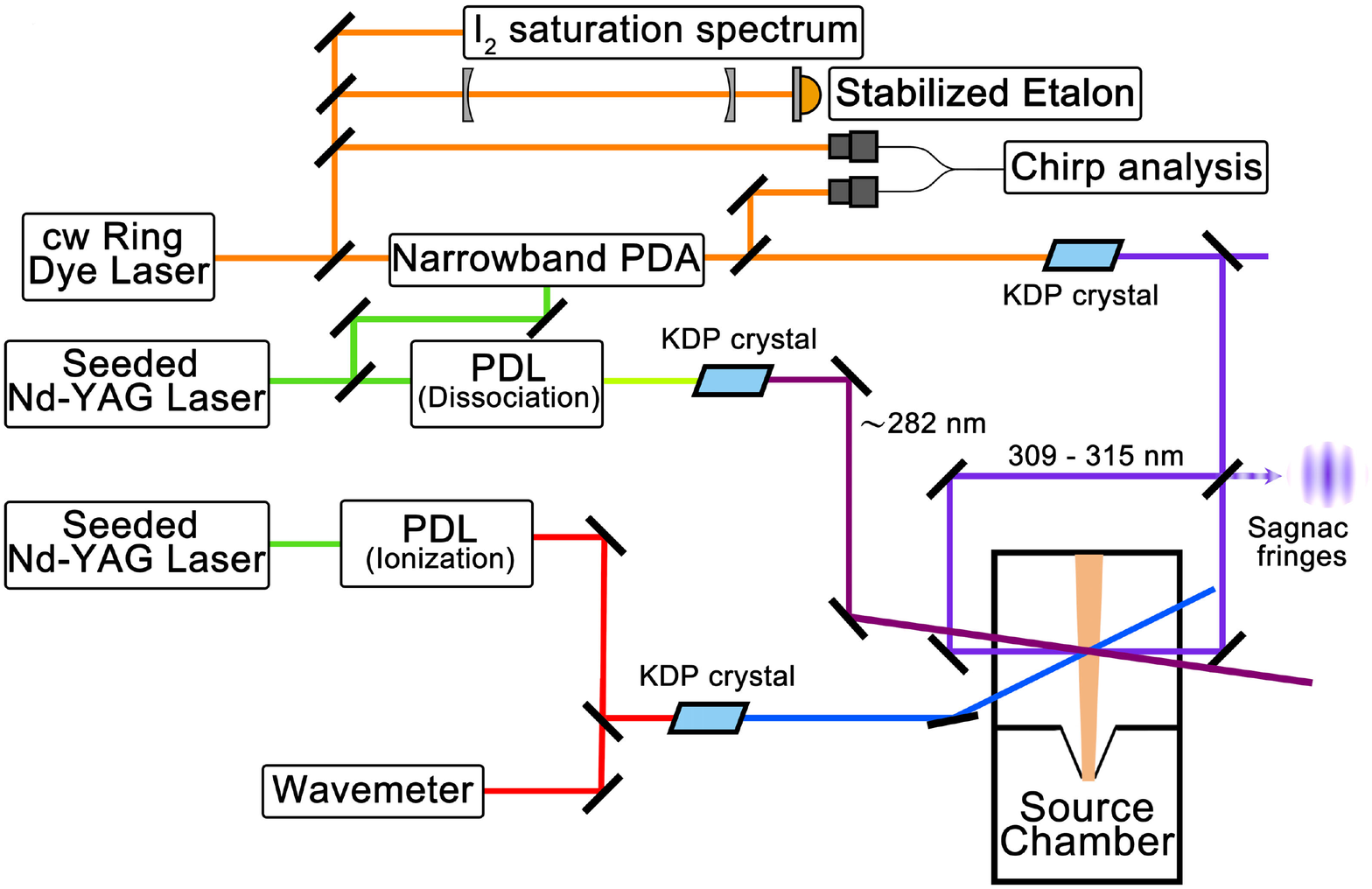}
\caption{\label{setup}
Schematic layout of the experimental setup including the three UV-lasers, the calibration units and the vacuum chambers. For details see text.}
\end{center}
\end{figure}

The experimental layout, shown in Fig.~\ref{setup}, is similar to the one used in previous studies probing \Hm\ in $v = 11, 12$~\cite{Trivikram2016,Trivikram2019}. Three ultraviolet (UV) pulsed laser systems are involved for producing the highest vibrational levels in \Hm\ from \hsm\ photolysis and detection by 2+1' REMPI. The two-photon UV-photolysis proceeds via the path:
\begin{equation*}
    \text{H$_{2}$S} \xrightarrow{2h\nu} \text{S} (^{1}{\rm D}_{2}) + \text{H$_{2}$} \text{(X $^1\Sigma_g^{+}$, $v=8-14$)}
\end{equation*}
The wavelength of the dissociation laser is set at 281.8~nm, as opposed to 291~nm in the previous studies, since the energy required for complete dissociation of \hsm\ to form an S($^1$D$_2$) atom and two H($^2$S) atoms is about 69935(25)~\wn~\cite{Zhou2020}. The two-photon energy for 281.8 nm dissociation lies about 1000 \wn\ above this limit, which is needed to produce \Hm\ in the highest vibrational levels close to the dissociation limit. Focused UV-pulses at energies of 4.5 mJ are used for the photolysis step.

Vibrationally excited \Hm\ is interrogated by a narrowband pulsed-dye-amplifier (PDA) system via probing of the \F-\X\ two-photon transition. The PDA is seeded by the output of a continuous wave (cw) ring-dye laser and its pulsed output is frequency doubled in a KDP crystal to deliver wavelength tunable  UV pulses in the range 309 - 315 nm. The bandwidth of this PDA is about 150 MHz in the UV. The UV pulse is split and configured into a counter-propagating geometry, shown in Fig.~\ref{setup}, and adjusted into a Sagnac interferometric alignment for reducing possible Doppler shifts~\cite{Hannemann2007}. The absolute frequency of the cw-seed light is calibrated by measurement of hyperfine-resolved saturation spectra of I$_2$ for reference, where markers of a stabilized etalon are used for interpolation. The chirp of the pulses of the PDA is analyzed and corrected for following known procedures~\cite{Eikema1997}.

\begin{figure}[b]
\begin{center}
\includegraphics[width=0.8\linewidth]{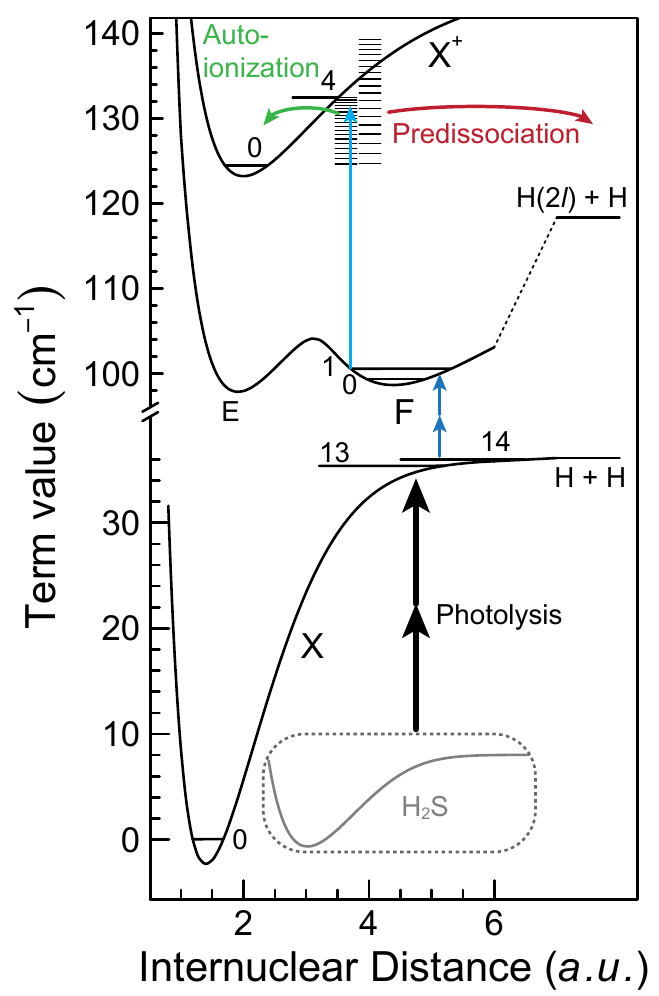}
\caption{\label{ex_scheme}
The excitation scheme followed in the present study.  The highly excited H$2$ ($v=13,14$) states are produced by two photon UV-photolysis of H$_2$S. These states are subsequently interrogated via 2+1' resonance-enhanced multi-photon ionization (three-laser scheme), while some overview spectra are recorded via via 2+1 REMPI (two-laser scheme).
For further details see text.}
\end{center}
\end{figure}

The third UV pulse, obtained from another frequency-doubled pulsed-dye-laser (LIOP-TEC), excites population in the \F-state into the  H$_2^{+}$ ionization continuum for detection. The autoionization spectra from F-states are recorded by scanning through 315 - 320 nm. The frequency of PDL output is calibrated with a HighFinesse WSU-30 (Toptica) wavemeter. In the case when precision measurements on the F-X transitions are performed, the third UV-laser is set on a strong autoionization resonance for signal optimization. Figure~\ref{ex_scheme} illustrates the level structure of the H$_2$ molecule and the various excitation steps induced in the three-laser scheme.

All three UV beams are focused to a few tens of $\mu$m and are spatially overlapped with the \hsm\ effusive molecular beam. The PDA-spectroscopy laser is optically delayed by 10 ns from the dissociation laser, which is pumped by the same Nd:YAG pump laser. This is to avoid an ac-Stark shift induced by the photolysis laser. For the same reason, the ionization laser is electronically delayed from the spectroscopy laser by 30 ns. H$_2^{+}$ ions produced are extracted into the mass-resolving time-of-flight tube and detected on an multichannel plate. The ion optics are triggered at about 50 ns delay from the ionization laser for avoiding dc-Stark fields during excitation.

\section{Results}

\begin{figure}
\begin{center}
\includegraphics[width=\linewidth]{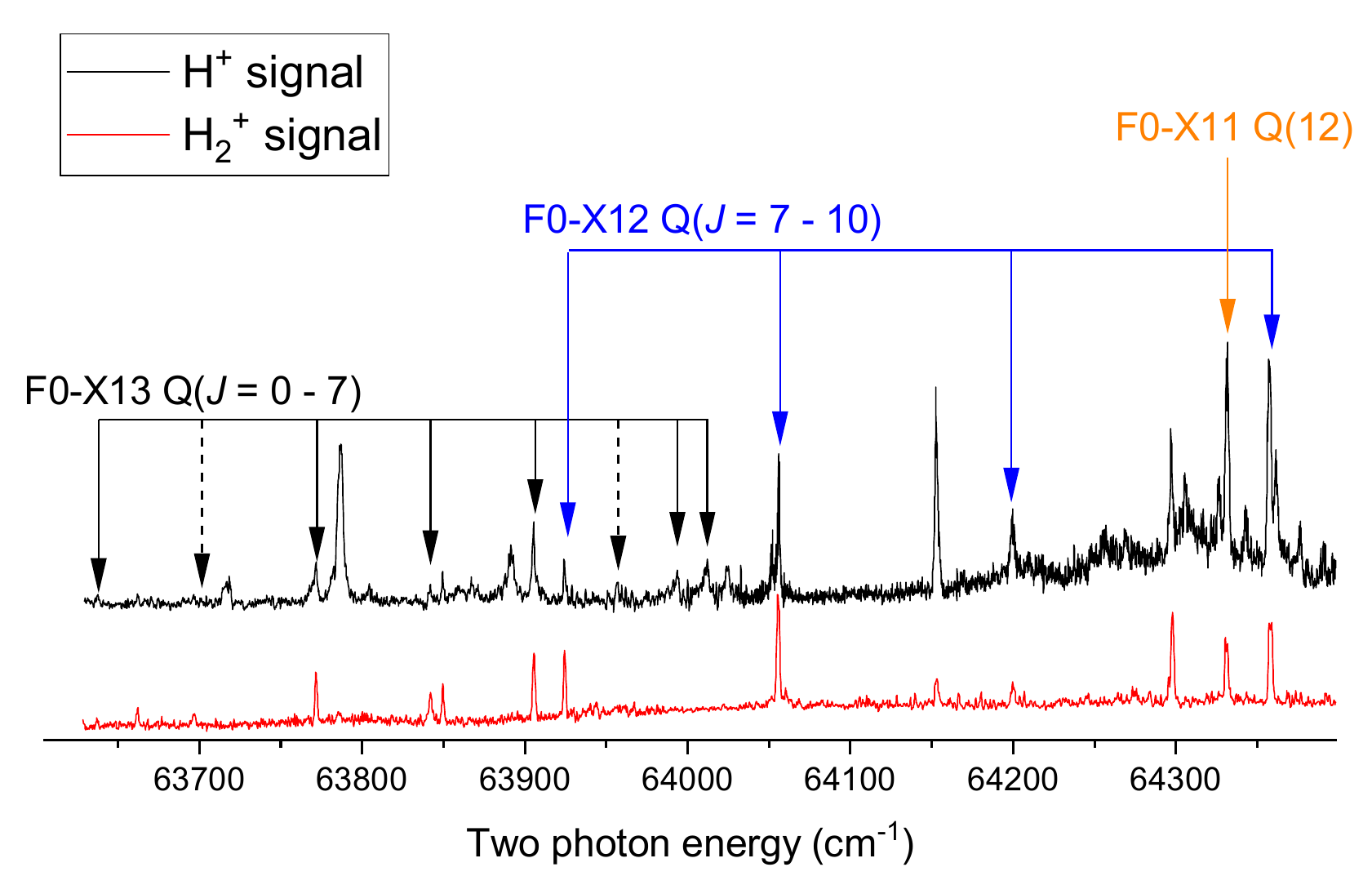}
\caption{\label{Overview}
Low-resolution overview spectra of the \F-\X(0,13) and \F-\X(0,12) bands probed via one-color 2+1 REMPI with a tunable frequency-doubled pulsed dye laser upon H$_2$ photolysis. Signals are recorded  for both H$^+$ and H$_2^{+}$.}
\end{center}
\end{figure}

First an overview spectrum was recorded using 2+1 REMPI on the \F-\X\ system in H$_2$ probing the population of high vibrational states in \X. This is done in a two-laser experiment, photolysis followed by one-color 2+1 REMPI, similar as in~\cite{Niu2015b}, using both H$_2^+$ and H$^+$ detection. Here the resolution is limited by the bandwidth of the frequency-doubled pulsed dye laser (\~ 0.1 \wn) used in the spectroscopy step. This overview spectrum, presented in Fig.~\ref{Overview},  displays rotational  Q-lines in the F0-X12 and F0-X13 bands, along with some additional resonances, some of which could not be assigned. It is noted that the intensity of the lines is affected by the excitation step into the autoionization continuum via the resonant photon energy. The assignment of the F-X resonances derives from a comparison with combination differences between experimental level energies in \F~\cite{Bailly2010} and those of \X, obtained from the precise \emph{ab initio} computations~\cite{SPECTRE2019}.

\begin{figure}
\begin{center}
\includegraphics[width=\linewidth]{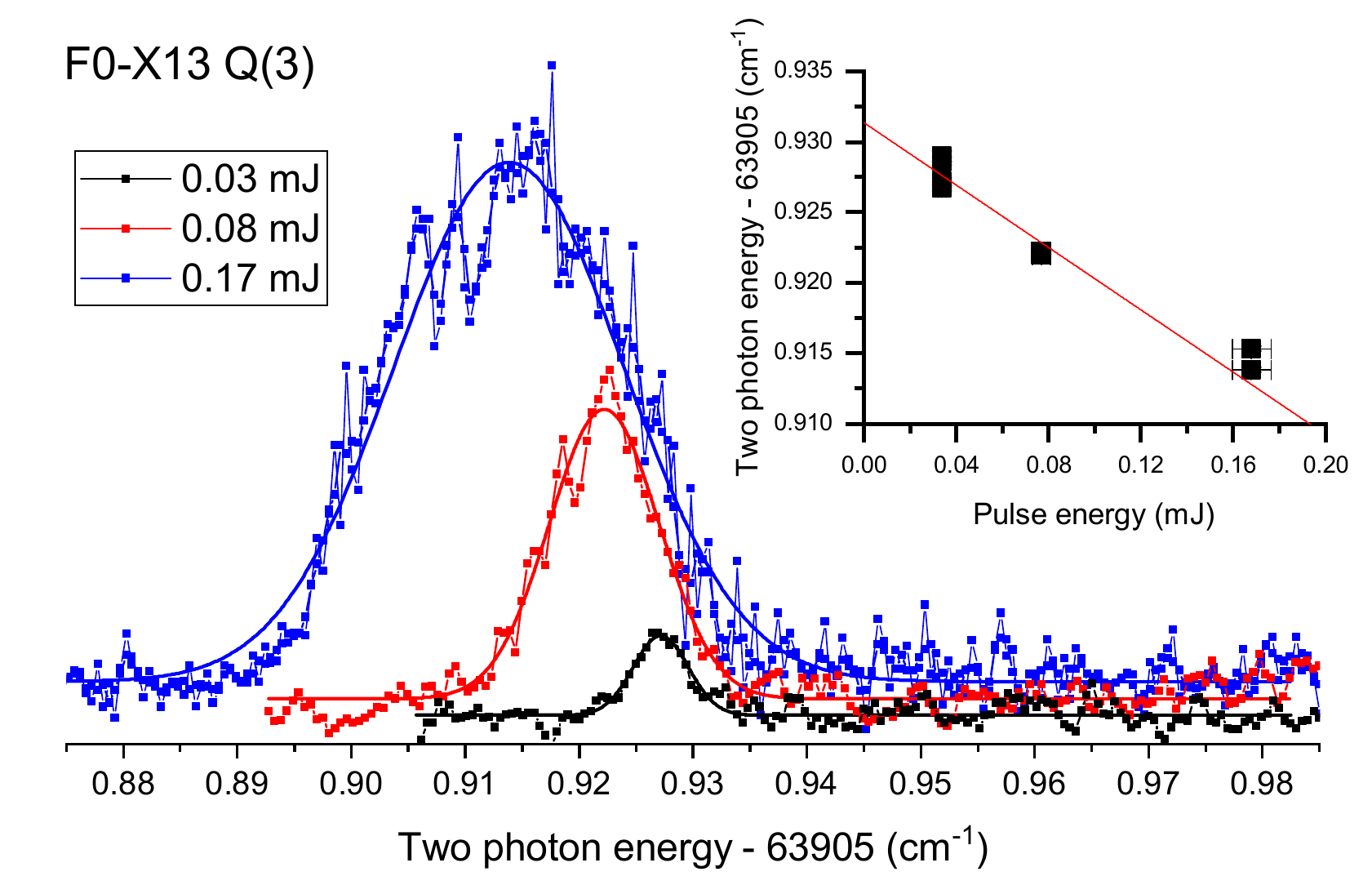}
\caption{\label{13q3}
Spectra of the F0 - X13 Q(3) transition recorded in a two-color 2+1' REMPI scheme with tuning of the narrowband frequency-doubled PDA-system with counter-propagating UV-beams. The inset shows the ac-Stark extrapolation to zero power levels.}
\end{center}
\end{figure}

\begin{figure}
\begin{center}
\includegraphics[width=\linewidth]{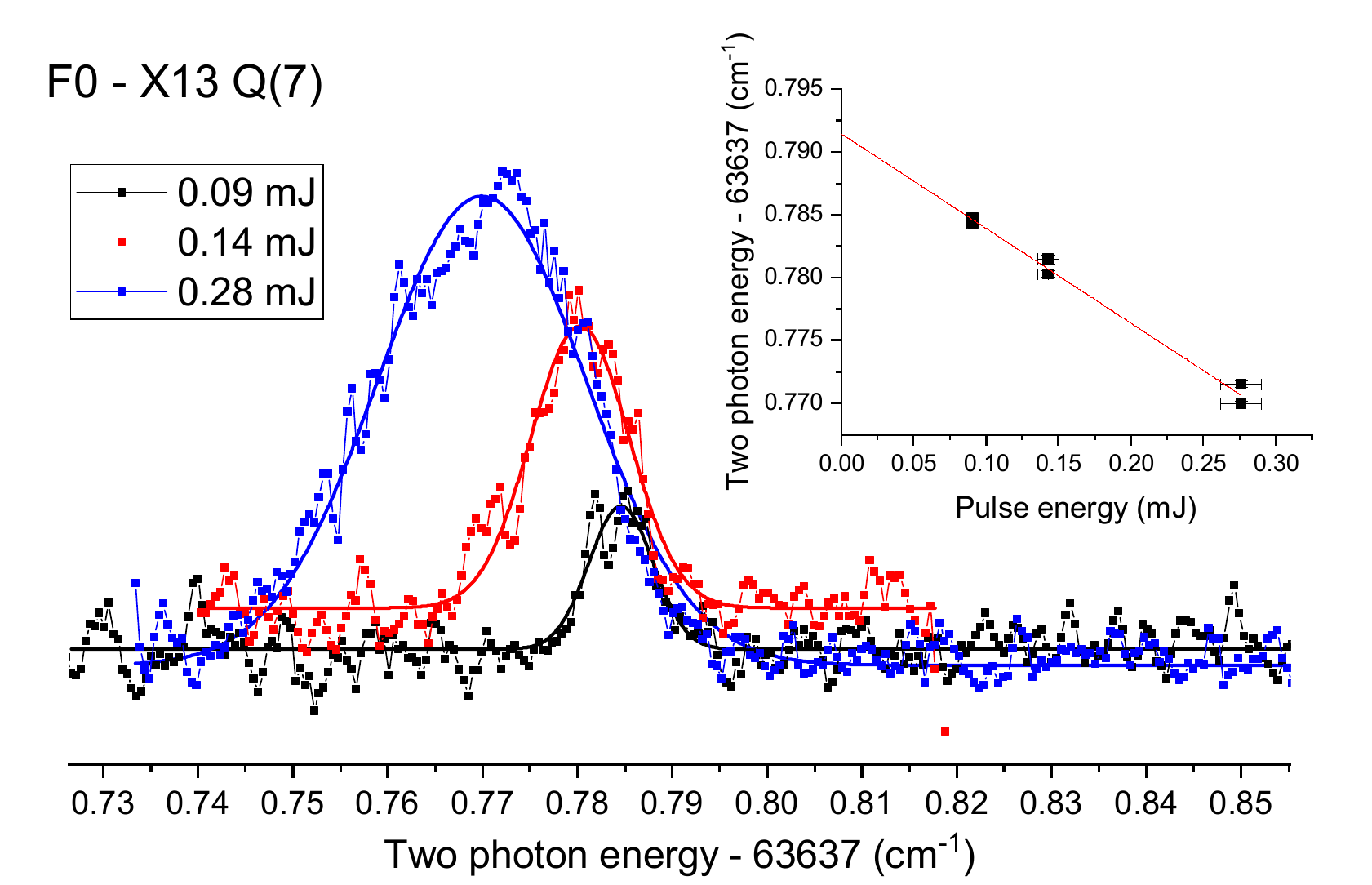}
\caption{\label{13q7}
Spectra of the F0 - X13 Q(7) transition; details as in Fig.~\ref{13q3}.}
\end{center}
\end{figure}

Subsequently, precision measurements of the \F-\X\ electronic transitions are recorded under Doppler-free conditions, applying  2+1' two-color REMPI in the three-laser scheme. While scanning the narrowband frequency-doubled PDA-system over the resonance, the third laser is set at a wavelength probing a strong as possible autoionization resonance, to be found in an iterative process. For these measurements H$_2^+$ ions are detected for registration of the spectra. Several Q-branch lines are measured probing  \X, $v = 13$ levels (denoted as X13) in excitation to the lowest vibrational level (F0) in the \F\ outer well for which the Franck-Condon factor is favorable~\cite{Fantz2006}.  All the odd $J$ states of $v = 13$ are detected, where $J = 7$ is the highest bound state for $v = 13$ in \Hm. Additionally, the Q(2) line in F0-X13 could be recorded, while the other even $J$ states of para-hydrogen appeared to be too lowly populated to be detected in the high resolution measurement.
Two of such spectra, for the Q(3) and Q(7) lines, are displayed in Figs.~\ref{13q3} and \ref{13q7}.

\begin{figure}
\begin{center}
\includegraphics[width=\linewidth]{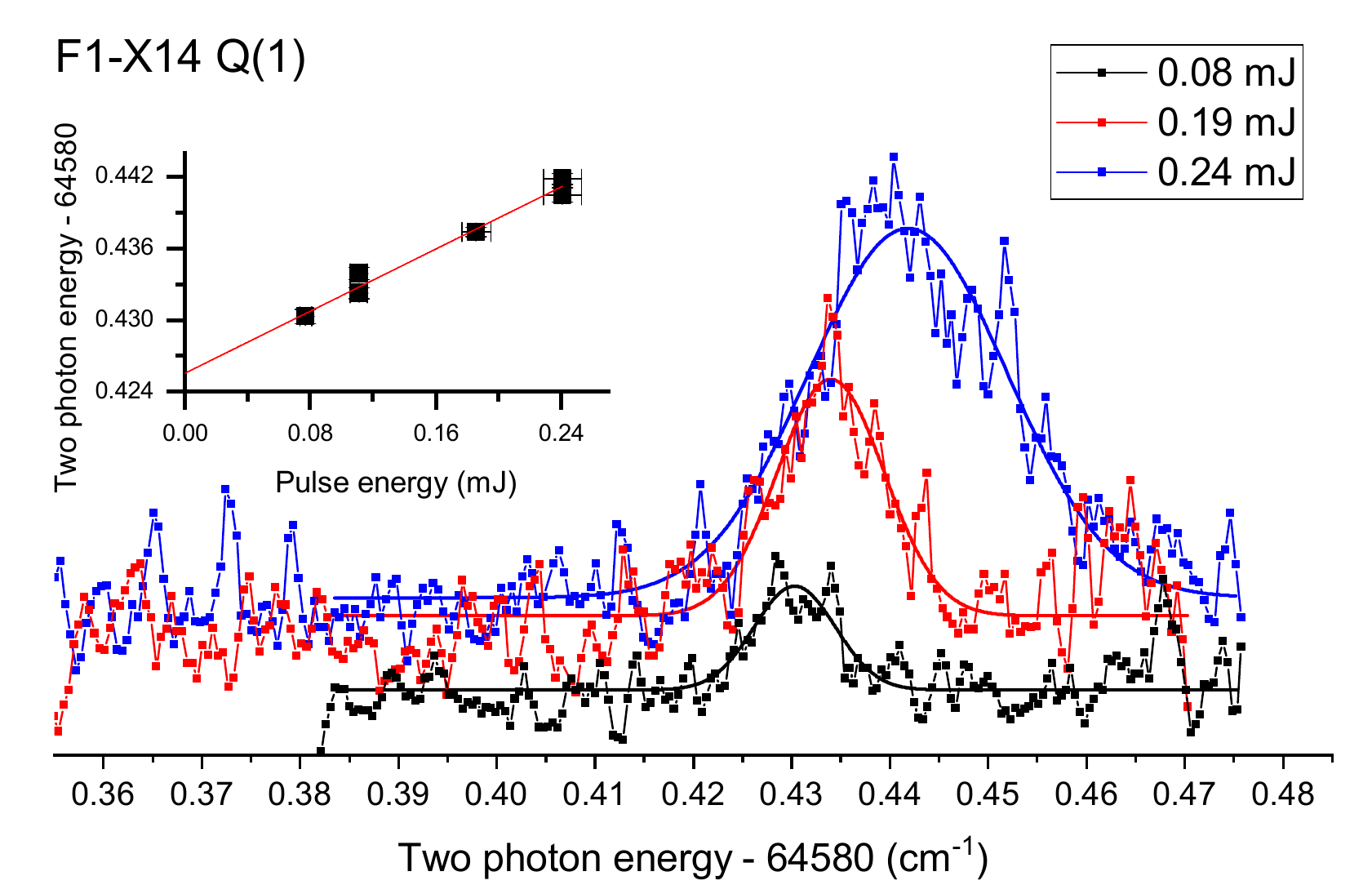}
\caption{\label{14q1}
Spectra of the F1 - X14 Q(1) transition; details as in Fig.~\ref{13q3}.}
\end{center}
\end{figure}

For the \X, $v = 14$ ground vibration  a high resolution recording could only be recorded for the $J=1$  level at a low signal-to-noise ratio. Power-dependent spectral recordings are shown in Fig.~\ref{14q1}, where also an ac-Stark extrapolation curve is displayed. Excitation from  other levels in $v=14$  remained below the noise level.

The various sources of measurement uncertainty for the F1-X13 Q-branch lines, recorded in the three-laser scheme, are listed in Table~\ref{tab:error}.
The major contribution to the total uncertainty is the statistical analysis over multiple sets of measurements, amounting to $2 \times 10^{-3}$~\wn.
The uncertainty in the frequency calibration of the cw-seed light, from the  measurement of I$_2$-hyperfine lines and interpolation of FSR-markers of the reference etalon, contributes overall $3 \times 10^{-4}$~\wn\ to the uncertainty.
The chirp-induced frequency offset between the pulse generated from PDA system and the cw-seed light has been analyzed through established techniques~\cite{Eikema1997},  adding $6 \times 10^{-4}$ \wn\ to the frequency uncertainties. For the latter two contributions a multiplication by a factor of four is included, for the frequency doubling and the two-photon process.
The Doppler-free two-photon excitation with counter-propagating beams, enforced by the Sagnac interferometric alignment~\cite{Hannemann2007}, constrains the uncertainty from a residual Doppler effect below $1 \times 10^{-4}$~\wn.
Since the ion optics are triggered at least 80~ns delayed from the spectroscopy laser  a dc-field-free environment is created, giving rise to a negligible dc-Stark effect to the accuracy.

The spectral recordings of the Doppler-free REMPI spectra undergo strong ac-Stark effects that contribute to the measurement uncertainty in two ways: asymmetric line profiles and shift of line centre.
The tightly focused PDA beams induce asymmetry of the line profile which limits the determination of line centre. The spectra are fitted with Gaussian and skewed Gaussian profile to account for this line profile asymmetry. The uncertainty estimated from the extrapolation to zero-power yields 0.0005 \wn.
For the low-power  spectrum of the F0-X13 Q(3) line, with 0.03 mJ UV pulse energy, the line width  is symmetric and about 180 MHz (FWHM), close to the expected instrumental linewidth determined by the laser bandwidth. For higher UV pulse energies, the spectral profiles become broadened and show a significant degree of asymmetry as result of spatial distribution of ac-Stark shifts in a tightly focused beam~\cite{Li1985}.
These lines are fitted with skewed Voigt profiles to determine the transition frequencies, as discussed previously~\cite{Trivikram2016}. The field-free transition frequencies are determined by extrapolation to zero power levels as shown in the insets of the figures.

\begin{table}
\caption{\label{tab:error}
Error budget for the two-photon frequencies for F0-X13 Q($J$).
}
\begin{tabular}{lc}
Contribution & Uncertainty ($\times 10^{-3}$ \wn\ ) \\
\hline
Line profile (fitting) & 0.5\\
Statistics & 2\\
AC-Stark extrapolation & 1 \\
Frequency calibration & 0.3\\
Cw-pulse offset (chirp)  & 0.6\\
Residual Doppler & $< 0.1$\\
DC-Stark effect & $< 0.1$\\
\hline
Total & 2.4\\
\end{tabular}
\end{table}

The overall uncertainty results in 0.0024 \wn, corresponding to 70 MHz,  by summing in quadrature for the F0-X13 Q-branch. For the F1-X14 Q(1) line 0.0040 \wn\ uncertainty is estimated, in view of the larger statistical uncertainty as a result of the poor signal-to-noise ratio obtained.
Transition frequencies determined for the observed F0-X13 Q($J$) and F1-X14 Q(1) lines, and the uncertainties, are listed in Table~\ref{tab:transition}.

\begin{table*}
\caption{\label{tab:transition}
Measured frequencies for the two-photon F-X transitions probing the highly excited vibrational levels \X, $v=13,14$, with uncertainties indicated in parentheses. These experimental values are compared with predicted values obtained via the combination of computed results for \X~\cite{SPECTRE2019,Czachorowski2018} and experimental values for \F~\cite{Bailly2010}.
The value for the F0 $J = 7$ level is absent in Ref.~\cite{Bailly2010}.
The uncertainties in the experimental F-state level energies were reevaluated from the data reported in the Supplementary Material of Ref.~\cite{Bailly2010} and listed under $\delta$F$_{\rm exp}$.
Similarly the uncertainties in the calculated values of the X-levels, as determined from the H2SPECTRE program~\cite{SPECTRE2019} are listed under $\delta$X$_{\rm theo}$. All values in \wn.
}
\begin{tabular}{ccccccc}
	&  & Exp. &  $\delta$F$_{\rm exp}$  &  $\delta$X$_{\rm theo}$ & Predicted   & Diff. \\
\hline
       & Q(1) & 63\,993.7920 (24) &  0.004 & 0.0035  & 63\,993.7956 (53)  & -0.0036 (58) \\
       & Q(2) & 63\,957.5160 (24) &  0.002 & 0.0034  & 63\,957.5236 (39)  & -0.0076 (46) \\
F0-X13 & Q(3) & 63\,905.9305 (24) &  0.010 & 0.0033  & 63\,905.9369 (105) & -0.0064 (108) \\
       & Q(5) & 63\,771.9836 (24) &  0.015 & 0.0028  & 63\,771.9723 (152) & +0.0113 (154) \\
       & Q(7) & 63\,637.7937 (24) &        &  0.0021  & \\
\hline
\hline
F1-X14 & Q(1) & 64\,580.4274 (40) &  0.009 & 0.0017  & 64\,580.4096 (92) & +0.0178 (95)\\
\hline
\end{tabular}
\end{table*}

\section{Discussion}

The dissociation of \hsm\ in the present study is performed via two-photon absorption at 281.8 nm instead of 291.5 nm as was used in the previous studies~\cite{Trivikram2016,Trivikram2019}, where the highest level observed in H$_2$ (\X) was $v = 12,  J = 5$. The energy required for complete dissociation of ground state \hsm\ into an S($^1{\rm D}_2$) atom and two H atoms is about 69935(25) \wn~\cite{Zhou2020}. The corresponding two-photon energy at 291.5 nm would reach only to about 1300 \wn\ below the \Hm\ dissociation limit, insufficient to produce  \Hm\ fragments in $v = 13$ and $v = 14$. A 2+1 REMPI spectrum of \hsm\ shows a resonance at 281.8 nm, and when fixing the dissociation laser wavelength to this resonance at  281.8 nm the energetic region of 1000 \wn\ above the dissociation limit of H$_2$ can be probed and $v = 13, 14$ produced. The signal strength for odd-$J$ transitions is generally stronger than for even-$J$ transitions, reflecting the ortho-para distribution which is apparently maintained in the photolysis process.
However, the signal intensity depends also on the efficiency of the autoionization induced by the setting of the third UV-laser.

In Table \ref{tab:transition} a comparison is made between measured transition frequencies  and predicted frequencies extracted from the calculated binding energies of \X ($v=13$) levels~\cite{SPECTRE2019,Czachorowski2018}  and \F-state level energies obtained from Fourier-Transform (FT) spectroscopic  measurements~\cite{Bailly2010}.
The uncertainty of the \X\ state binding energies as obtained from the H2SPECTRE program suit~\cite{SPECTRE2019} amounts nominally to 0.002 - 0.003 \wn\ with computed values listed in Table~\ref{tab:transition}.
For the experimentally determined level energies of F0, obtained up to $J=5$ only, we have reanalyzed the uncertainties, based on the data presented in the Supplementary Material of Ref.~\cite{Bailly2010}, because precision studies on the GK-X transition in \Hm~\cite{Cheng2018} have shown that some levels in Ref.~\cite{Bailly2010} exhibit a somewhat larger uncertainty than previously estimated. The reevaluated uncertainties are listed in Table~\ref{tab:transition}.
This comparison leads to fair agreement.

\begin{figure*}
\begin{center}
\includegraphics[width=\linewidth]{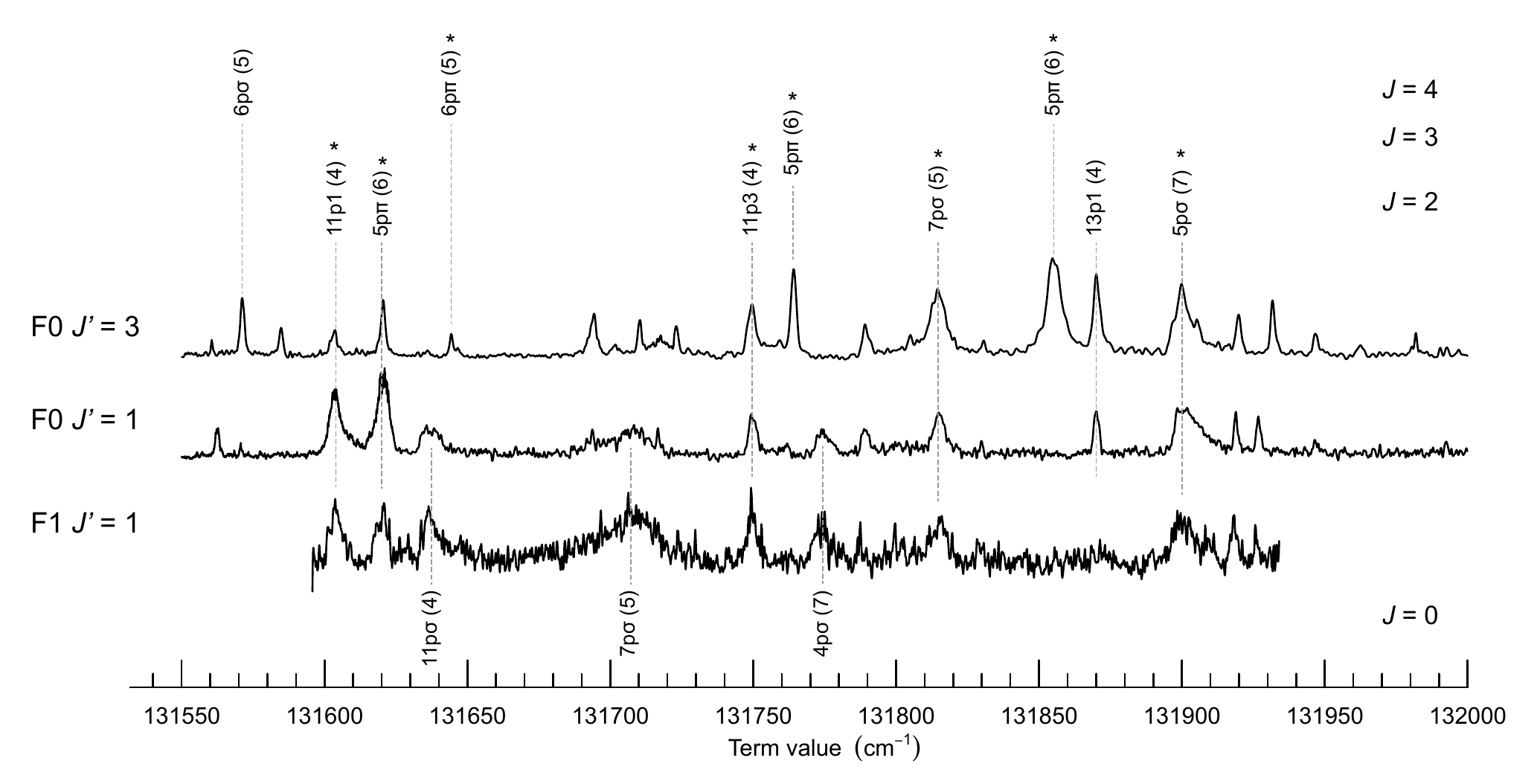}
\caption{\label{autoion}
Autoionization spectra from F0 $J' = 1$, F1 $J' = 1$ and F0 $J' = 3$ intermediate levels. The horizontal axis reflects the total term value by adding the UV two-photon energy and the F-state level energy from Ref.~\cite{Bailly2010}: 99376.0474 \wn\ for F0 $J = 1$, 100570.843 \wn\ for F1 $J = 1$ and 99437.1665 \wn\ for F0 $J = 3$. The resonances in the autoionization continuum are labeled in either Hund's case b) - $n$p$\lambda(v^+)$ - or d) -$n$p$N^+(v^+)$ -, where $v^+$ and $N^+$ are the vibrational and the rotational quantum numbers of the H$_2^+$ ion core. The total angular momentum quantum number $J$ of the H$_2$ continuum resonance is given on the right. Previously observed resonances \cite{Glass-Maujean2013a,Glass-Maujean2013b,Glass-Maujean2013c} are indicated by an asterisk.}
\end{center}
\end{figure*}

\begin{table*}[t]
\renewcommand{\arraystretch}{1.3}
\caption{\label{tab:com_diff}
Combination difference with measured F0-X11 transitions in Ref.~\cite{Trivikram2016} and comparison with calculations in the ground electronic state. All values are presented in \wn, with uncertainties indicated in parentheses.}
\begin{tabular}{rccccc}
     & F0-X13         & F0-X11         & X13-X11       & Calculation          & Difference     \\
\hline
Q(1) & 63\,993.7920 (24) & 66\,438.2920 (15) & 2\,444.5000 (28) & 2\,444.5019 (53) & -0.0019 (60) \\
Q(3) & 63\,905.9305 (24) & 66\,250.6874 (15) & 2\,344.7569 (28) & 2\,344.7489 (52) & +0.0080(59)  \\
Q(5) & 63\,771.9836 (24) & 65\,931.3315 (15) & 2\,159.3479 (28) & 2\,159.3459 (51) & +0.0021(58)  \\
Q(7) & 63\,637.7937 (24) & 65\,510.6124 (20) & 1\,872.8187 (31) & 1\,872.8277 (48) & -0.0090 (57) \\
\hline
\end{tabular}
\end{table*}

In order to test the full QED-relativistic calculations of the ground state binding energies, a further comparison is made using combination differences from experiment, independent of data on F-level energies from Ref.~\cite{Bailly2010}.
The previous experimental values on F0-X11 Q-transitions~\cite{Trivikram2016} are subtracted from the present values for F0-X13 Q-transitions to obtain vibrational splittings between X13 and X11 levels which can be compared with the same combination differences from the most advanced first principles calculations~\cite{SPECTRE2019,Czachorowski2018}.
Table \ref{tab:com_diff} presents these comparisons for four sets of $J$-levels. The experimentally derived splittings between $v = 11$ and $v = 13$  agree with the calculations at a root-mean-square  value of $\pm 1.07 \sigma$.

These result provide a test on the accuracy of the ab initio computations for ground state levels, for the first time for the highest $v$-levels in the \X\ ground state of H$_2$.
The current theoretical values are limited by non-adiabatic contributions to the relativistic energy~\cite{Czachorowski2018}. Meanwhile, improved fully variational calculations have been developed, that led to an uncertainty less than $10^{-7}$~\wn, but as of yet only for the \X\ ($v=0, J=0$) ground state~\cite{Komasa2019}.

As for the F1-X14 Q(1) line no such comparison could be made, because the F1-X13 and F1-X12 Q(1) transitions appeared too weak in the high precision measurement.
In the search of $v = 14$ in \Hm, excitations to F0, F1 and F2 states were tested but only a single F1-X14 Q(1) line is confirmed. The absence of F0-X14 lines could be explained by the small Franck-Condon factor, which is about 70-times smaller than that for F1-X14~\cite{Fantz2006}.
The assignment of F1-X14 Q(1) is verified by comparing autoionization spectra recorded in the region below the X$^+(v^+=4,J^+=1)$ ionization threshold as shown in Fig.~\ref{autoion}.
The F-outer-well state has nominally $(2\text{p}\sigma_u)^2$ character and we observe exclusively transitions to vibrationally autoionizing $(1\text{s}\sigma)(n\text{p}\sigma/\pi)$ Rydberg states, with $n$ being the principal quantum number. To guide the assignment we carried out multichannel quantum-defect (MQDT) calculations as described in Ref.~\cite{Jungen1977} using the quantum-defect functions derived in Ref.~\cite{Sprecher2014}. We note that states with $\Sigma^+$ and $\Pi^+$ symmetry are subject to predissociation into the $\text{H}+\text{H}(n=2)$ continuum, leading to broad resonances especially for low $n$-values. The interaction with the dissociation continuum was not included in our calculation, leading to deviations of the calculated term values for the low-$n$ resonances on the order of 1~cm$^{-1}$. The experimental line positions for these states are however in good agreement with previously reported values and MQDT calculations including the combined ionization and dissociation continuum \cite{Glass-Maujean2013a,Glass-Maujean2013b,Glass-Maujean2013c}.

Autoionization spectra are recorded by fixing the PDA-spectroscopy laser on two-photon resonances probing F0 $J=1$, F0 $J=3$ and F1 $J=1$, and are plotted on an energy scale relative to the X, $v = 0, J = 0$ state of \Hm.
The fact that the autoionization spectrum from F0 $J=1$, probed via X13 $J=1$, exhibits the same resonances as the autoionization spectrum from F1 $J=1$, probed via X14 $J=1$, proves that the two-photon resonance at $64\,580.427$ \wn\ starts from a $J=1$ line, which leads to an unambiguous assignment of the F1-X14 Q(1) transition.
Unfortunately no other rotational levels in X14 could be found. There appears a strong transition at 64563.084 \wn, while the expected F1-X14 Q(3) is at 64562.883 \wn, exhibiting a 0.201 \wn\ difference. Also in this case an autoionization spectrum is recorded from this intermediate state, but that does not match with an autoionization spectrum from F0 $J = 3$. Hence the assignment of the F1-X14 Q(3) line is discarded.

As a byproduct of the present study the Stark slopes of the two-photon transitions are determined, results of which are shown in the insets of Figs.~\ref{13q3}-\ref{14q1}. Those represent the shift of line center as a result of the ac-Stark effect, i.e. the power density; a negative Stark slope corresponds to a red-shift of the lines for higher power densities. Stark slopes for the F0-X13 band are all negative, while that of the single transition in the F1-X14 band is found positive.  In a previous study a negative Stark slope was found for lines in the F0-X11 band, where a very small $J$-dependent value was found for the F3-X12 band~\cite{Trivikram2016}. Similarly, in two-photon excitation to the inner well positive Stark slopes were found for the E0-X0 band~\cite{Hannemann2006} and the E0-X1 band~\cite{Niu2014}. These Stark slopes depend on the transition dipole moments in summation over all states in the molecule~\cite{Girard1983,Huo1985} thus providing information on the quantum structure of the molecule, analysis of which is beyond the scope of this study.

\section{Conclusion}

In the present study two-photon UV-photolysis of \hsm\ was pursued, probing a two-photon absorption resonance at 281.8 nm, hence  at sufficiently short wavelength to produce \Hm\ molecules in the highest vibrational levels: $v=13,14$.
The transition energies of F-X (0,13) Q-branch lines have been measured at an accuracy of 0.0024~\wn. In comparing these results with previous measurements on F0-X11 Q-lines the experimental combination difference can be used to verify level splittings as computed with advanced quantum chemical calculations of the ground electronic state of \Hm, including non-adiabatic, relativistic and QED effects. This results in good agreement.
Also the highest vibrational state $v = 14$ of \Hm\ has been produced through the two-photon photodissociation of \hsm. Only a single rotational level $J=1$ could be observed. The assignment of this \X\ ($v=14, J=1$)  level was verified by recording and comparing autoionization spectra from various F-outer well states.

The total uncertainty in the present study is limited by measurement statistics, associated with the low concentration of \Hm\ fragments produced, and by strong ac-Stark effects resulting from the focused UV-laser beams required to obtain signal. Further improvement of the \hsm\ photolysis production process is critical for pursuing higher measurement accuracy. Under such improved conditions the entire rotational manifold $J=0-3$ of bound levels in $v=14$ might be probed and studied and implemented in QED-test of such weakly bound states at large internuclear separation.
It has been a matter of debate whether the final level $J=4$ is rotationally predissociative~\cite{Leroy1971}, only bound by non-adiabatic effects~\cite{Pachucki2009}, or quasi-bound due to hyperfine effects~\cite{Selg2011}.
As an outlook it may be hypothesized that also
quasi-bound states in \Hm\ might be observed by the methods pursued here. 

\section*{Acknowledgement}

The authors thank Dr. Christian Jungen for fruitful discussions and for making available his MQDT-codes for calculating and assigning the autoionization resonances. WU acknowledges the European Research Council for an ERC Advanced grant (No: 670168).

%

\end{document}